\documentclass[aps,prl,reprint,superscriptaddress]{revtex4-2}

\usepackage{graphicx}
\usepackage{dcolumn}
\usepackage{bm}

\begin{document}

\preprint{APS/123-QED}

\title{Wake-Powered Spreading of Encapsulated Drops Impacting Under Liquid}

\author{Tian-Yu Zhang}
 \affiliation{Micro \& Nano-scale Transport Laboratory, Waterloo Institute for Nanotechnology, Department of Mechanical and Mechatronics Engineering, University of Waterloo, Waterloo, Ontario N2L 3G1, Canada}
 
\author{Sushanta K. Mitra}%
 \email{contact author: skmitra@uwaterloo.ca}
 \affiliation{Micro \& Nano-scale Transport Laboratory, Waterloo Institute for Nanotechnology, Department of Mechanical and Mechatronics Engineering, University of Waterloo, Waterloo, Ontario N2L 3G1, Canada}
\affiliation{
 International School of Engineering, Faculty of Engineering, Chulalongkorn University, Bangkok 10330, Thailand\\
}%

\date{\today}

\begin{abstract}
A drop impacting a wall in air spreads by converting its own kinetic energy into surface energy. We show that this picture fails when the wall is submerged in liquid. Millimetric perfluoropolyether drops, bare or encapsulated in an ultrathin silicone-oil shell, impacting a glass wall in water are flattened by the squeeze flow in the draining gap and shed most of their kinetic energy before impact. Nevertheless, they spread to more than twice their diameter on a water lubrication film that never ruptures, creating more surface energy than their entire kinetic and gravitational energy could supply. A two-stage energy scaling resolves this deficit. The liquid set in motion in the drop’s wake supplies more than 90\% of the energy for spreading, so that the maximum spreading is governed by the Weber number of the ambient liquid rather than that of the drop. The shell acts mainly through its interfacial tension, while its biaxial-extension dissipation, negligible before impact, becomes appreciable during spreading.
\end{abstract}

\maketitle

\textit{Introduction}---When a drop hits a solid wall in air, its fate is set by a closed energy budget: the kinetic energy carried by the drop is converted into surface energy and viscous dissipation, which fixes the maximum spreading factor ${\beta }_{\mathrm{m}}={D}_{\mathrm{m}}/{D }_{\mathrm{eq}}$ \cite{1,2,4,5,6}. The surrounding gas is a passive spectator except in the thin films that control splashing, skating, and bouncing \cite{7,8,9}. In many processes, however, the ambient fluid is a liquid of density and viscosity comparable to those of the drop, such as immersion cooling \cite{10,11}, bioprinting \cite{12}, underwater self-cleaning \cite{13}, and oil-water separation \cite{14,15}. The ambient liquid must then be squeezed out of the gap before the drop can reach the wall, a drainage lasting milliseconds rather than microseconds \cite{16,17}, and it is itself set in motion by the falling drop. How these two effects reorganize the spreading behavior and energy budget of impact is unknown. Studies of drops meeting immersed walls have addressed trajectories and rebound \cite{18} or wettability \cite{15,19}. Those on liquid-liquid interfaces have addressed coalescence and wake interaction \cite{20,21}, while quasi-static spreading under liquid is governed by contact-line physics \cite{22,23,24}.

The question is sharpened for compound drops, in which a liquid core is wrapped by a thin immiscible shell \cite{25,26,27}. In air, the shell adds an interface and a viscous layer that alter spreading and rebound \cite{28,29,30,31}. Liquid-liquid encapsulation (LLE) \cite{32,33,35,36,37} produces such drops entirely under liquid, where they inevitably impact submerged walls, yet their underliquid impact has not been examined.

In this Letter, we show that underliquid impact is an open-system process in which the ambient liquid first removes and then resupplies the energy of the drop. A squeeze flow flattens and decelerates the drop before its front stops, and the drop then spreads on a persistent water film, creating more surface energy than its own mechanical energy can provide. A two-stage energy scaling demonstrates that the deficit is paid by the wake flow. The maximum spreading is controlled by the Weber number of ambient liquid, and a thin viscous shell becomes dissipative only once spreading stretches it.

\textit{Experiments}---Drops of Krytox 1506 or 1514 (density ${\rho }_{\mathrm{core}}\approx 1.89\times {10}^{3}\,kg\,m^{-3}$, viscosity ${\mu}_{\mathrm{core}} = 113$ or $265\,mPa\,s$) were encapsulated in silicone oil AP1000 (${\mu }_{\mathrm{il}}=1000\,mPa\,s$) by constrained gravity-driven LLE \cite{36,37}. The core liquid was dispensed onto a $400\,\mu{L}$ silicone-oil layer held at the water bath surface by a floating hydrophobic ring, penetrated it, and pinched off wrapped in a shell of thickness ${\delta}_{0}$ of tens of micrometers [Supplemental Material (SM) \cite{38}, Fig. S1]. The drop then fell through the bath of depth $h_{\mathrm{bath}}=52-76\,mm$ onto a submerged quartz wall. Bare drops, produced identically without the shell liquid, served as reference. The equivalent diameter $D_{\mathrm{eq}}=4.1-5.5\,mm$ exceeds the underwater capillary length, giving Bond numbers $Bo=\Delta \rho gD_{\mathrm{eq}}^{2}/\Gamma=3.8-4.8$, where $\Delta \rho ={\rho }_{\mathrm{core}}-{\rho }_{\mathrm{bath}}$ and $\Gamma ={\gamma }_{\mathrm{core},\mathrm{il}}+{\gamma }_{\mathrm{il},\mathrm{bath}}\approx 39\,mN\,m^{-1}$ for encapsulated drops ($\Gamma ={\gamma }_{\mathrm{core},\mathrm{bath}}\approx 51-54\,mN\,m^{-1}$ for bare drops). As the unwrapping free energy ${\gamma }_{\mathrm{core},\mathrm{bath}}-\Gamma \approx 12-15\,mN\,m^{-1}$ is positive, the shell is thermodynamically retained during impact [35]. The drop terminal velocity $U_{0}=0.18-0.22\,m\,s^{-1}$ gives Weber numbers $We_{0}={\rho }_{\mathrm{core}}U_{0}^{2}D_{\mathrm{eq}}/\Gamma = 6-9$ and bath Reynolds numbers ${\rho }_{\mathrm{bath}}U_{0}D_{\mathrm{eq}}/{\mu }_{\mathrm{bath}}\approx {10}^{3}$. Impacts were captured at 2000 fps on walls ranging from underwater superoleophobic (SOPB) to oleophilic (OPL), each condition being repeated three times (SM, Secs. S1-S4).

\begin{figure*}
    \centering
    \includegraphics[width=\textwidth]{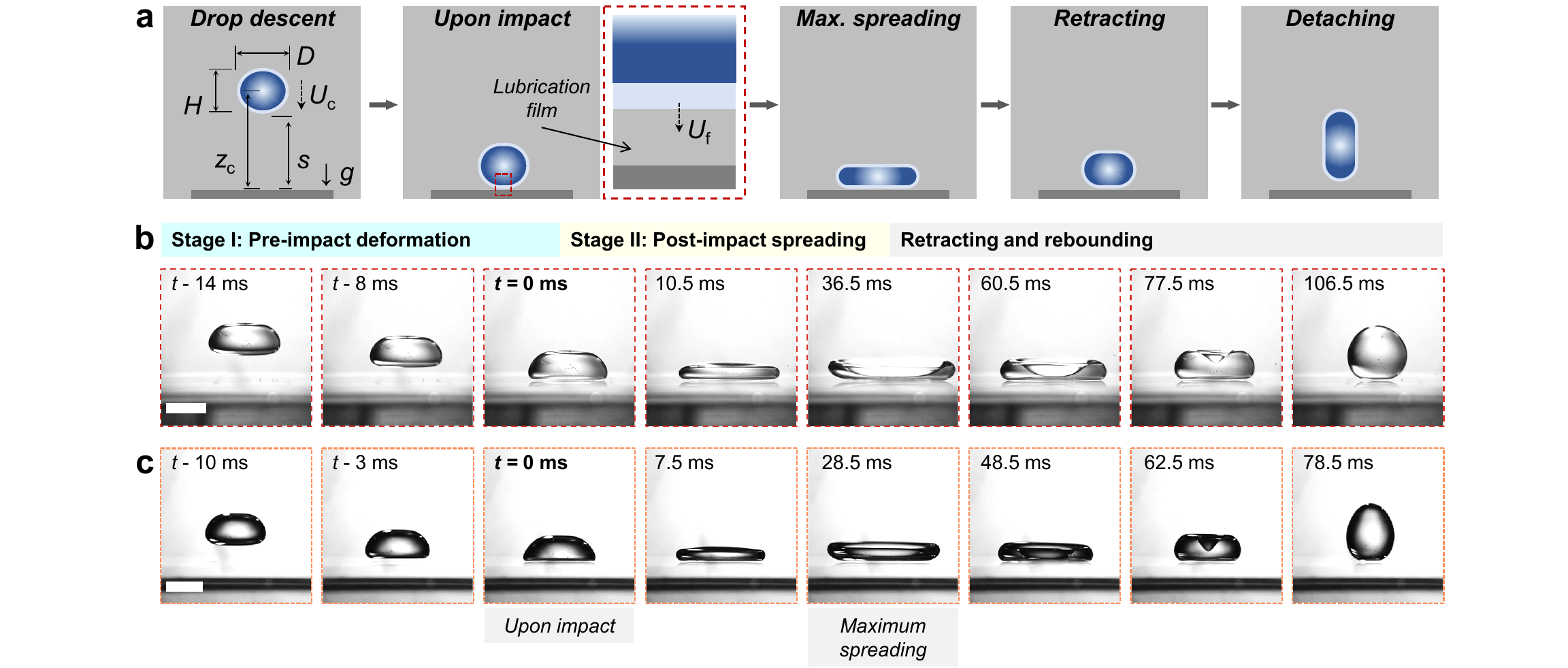}
    \caption{\label{fig1} Underliquid impact on a submerged wall. (a) Descent, impact on a water lubrication film (inset), maximum spreading, retraction, and detachment. $D$ and $H$ are the horizontal and vertical extents of the drop, $z_{\mathrm{c}}$ and $s$ the heights of its centroid and lower front above the wall, and $U_{\mathrm{c}}$ and $U_{\mathrm{f}}$ their velocities. (b) Bare Krytox 1514 drop ($D_{\mathrm{eq}}\approx 5.46\,mm$, $h_{\mathrm{bath}}=64\,mm$) and (c) Krytox 1506 drop encapsulated in AP1000 ($D_{\mathrm{eq}}\approx 4.16\,mm$, $h_{\mathrm{bath}}=68\,mm$) on the SOPB wall (Movies S1 and S2 \cite{38}). The lower front stops at $t=0$. Scale bars: 5 mm.}
\end{figure*}

\textit{Two stages of impact}---Figures 1 and 2 show that underliquid impact differs from its airborne counterpart in two aspects. First, the drop deforms before it arrives [Figs. 1(b) and 1(c)]. It is already oblate during descent ($D/D_{\mathrm{eq}}\approx 1.2$), and about $17.5\,ms$ before impact its lower front begins to lag the centroid [Fig. 2(a)]. The water squeezed out of the narrowing gap decelerates the front faster than the bulk, and the drop then flattens to $D/D_{\mathrm{eq}}\approx 1.45$ [Figs. 2(c) and 2(d)]. We therefore define impact kinematically, as the instant $t=0$ at which $U_{\mathrm{f}}<{10}^{-3}U_{\mathrm{c}}$ [Fig. 2(a)]. At that instant the centroid still moves at about half its terminal velocity, so the drop has already shed 75\%-95\% of its kinetic energy. This pre-impact deformation (stage I) supplies 20\%-30\% of the total lateral expansion. Second, the drop never wets the wall. During post-impact spreading (stage II) it flattens into a rimmed pancake, reaching ${\beta }_{\mathrm{m}}\approx 2.0-2.5$ after $30-40\,ms$, then retracts and lifts off without contact-line pinning, even on the OPL wall (SM, Sec. S5). Spreading thus proceeds on a persistent water lubrication film, the underliquid counterpart of air-mediated bouncing \cite{8,9}, and is only weakly sensitive to surface wettability. Encapsulated drops follow a similar sequence [Figs. 1(c) and 2(d)].

These observations pose a paradox. During impact spreading, the drop can draw on its kinetic energy $K_{0}=\pi {\rho }_{\mathrm{core}}D_{\mathrm{eq}}^{3}U_{0}^{2}/12$ and on the gravitational work of its centroid descent, which together amount to $1.8-4.6\,\mu{J}$. The surface energy it creates, $\Gamma \Delta A$, is $2.4-9.6\,\mu{J}$, a factor 1.3-2.1 larger, even if no viscous damping is counted (SM, Sec. S9). The drop must therefore draw energy from the surrounding liquid. To identify the source, we establish an energy balance for each stage, approximating the drop as a volume-conserving cylinder of width $D=\beta D_{\mathrm{eq}}$ and height $H=2D_{\mathrm{eq}}/\left(3{\beta }^{2}\right)$, with surface area $\frac{1}{2}\pi D_{\mathrm{eq}}^{2}S\left(\beta \right)$, where $S\left(\beta \right)={\beta }^{2}+4/\left(3\beta \right)$.

\begin{figure*}
    \includegraphics[width=\textwidth]{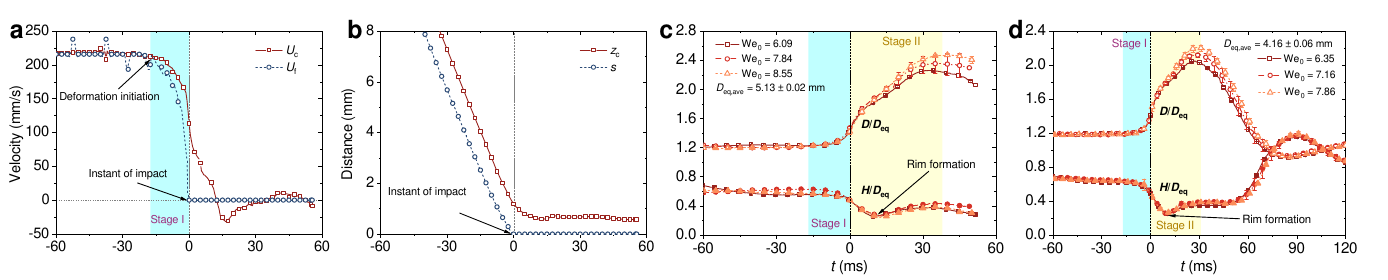}
    \caption{\label{fig2} Hydrodynamics of underliquid impact on the SOPB wall. (a) Centroid and lower front velocities (downward positive) and (b) centroid height $z_{\mathrm{c}}$ and gap $s$ for a bare Krytox 1506 drop ($We_{0}=8.55$). (c,d) Normalized width $D/D_{\mathrm{eq}}$ and height $H/D_{\mathrm{eq}}$ of (c) bare and (d) AP1000-encapsulated Krytox 1506 drops at three $We_{0}$. Shading marks stages I and II. Error bars are standard deviations of three trials.}
\end{figure*}

\textit{Stage I: squeeze-flow pre-deformation}---During stage I [Figs. 3(a)-3(e)], the lower front decelerates from $U_{0}$ to rest over the initial gap $s_{0}=\lambda D_{\mathrm{eq}}$, while the centroid travels a further $\left(H_{0}-H_{\mathrm{i}}\right)/2$ and decelerates to $U_{\mathrm{i}}$. Taking mean velocities $U_{0}/2$ and $\left(U_{0}+U_{\mathrm{i}}\right)/2$ gives the duration and the velocity ratio,
\begin{equation}
  t_{\mathrm{d}}\simeq \frac{2s_{0}}{U_{0}},
  \label{td}
\end{equation}
\begin{equation}
  \chi \equiv \frac{U_{\mathrm{i}}}{U_{0}}\simeq \frac{{\beta }_{0}^{-2}-{\beta }_{\mathrm{i}}^{-2}}{3\lambda },
  \label{chi}
\end{equation}
where ${\beta }_{0}\approx 1.2$ and ${\beta }_{\mathrm{i}}$ are the deformation factors at onset and impact. With the measured $\lambda \approx 0.35-0.38$, Eq.~(\ref{td}) predicts $t_{\mathrm{d}}=16-18\,ms$, in agreement with the observed $17.5\,ms$. Energy conservation from $t=-t_{\mathrm{d}}$ to 0 reads, 
\begin{equation}
  K_{0}+W_{\mathrm{g},\mathrm{I}}=K_{\mathrm{i}}+\Delta E_{\Gamma ,\mathrm{I}}+E_{\mathrm{sq}}+E_{\mathrm{core},\mathrm{I}}+E_{\mathrm{il},\mathrm{I}}.
  \label{E_stageI}
\end{equation}

The squeeze flow dissipates a fraction $C_{\mathrm{sq}}$ of the work done by the hydrodynamic force $F_{\mathrm{hyd}}\simeq \Delta \rho Vg+mU_{0}\left(1-\chi \right)/t_{\mathrm{d}}$ on the decelerating front. The core dissipates through radial shear across its height. The ultrathin shell is stretched in uniform biaxial extension at constant volume $V_{\mathrm{il}}=A\delta$, and dissipates \cite{49}
\begin{equation}
  E_{\mathrm{il}}\simeq 3{\mu }_{\mathrm{il}}V_{\mathrm{il}}\int{\left(\frac{\dot{A}}{A}\right)}^{2}\mathrm{d}t\simeq \frac{3{\mu }_{\mathrm{il}}V_{\mathrm{il}}}{\tau }\ln^{2}\frac{A_{2}}{A_{1}},
  \label{EilI}
\end{equation}
over a stage of duration $\tau$ in which its area grows from $A_{1}$ to $A_{2}$. Normalizing by $\frac{1}{2}\pi \Gamma D_{\mathrm{eq}}^{2}$ yields (SM, Sec. S6),
\begin{eqnarray}
S(\beta_{\mathrm{i}})-S(\beta_0)
&=& \frac{We_0}{6}
\left[1-\chi^2-C_{\mathrm{sq}}(1-\chi)\right]
\nonumber\\
&+& \frac{Bo\lambda}{3}
\left(1-C_{\mathrm{sq}}+\chi\right)
-\mathcal{D}_{\mathrm{I}},
\label{stageI}
\end{eqnarray}
where Eq.~(\ref{td}) and (\ref{chi}) has been used, $Ca=\mu U_{0}/\Gamma$, and the core and shell dissipations are collected in ${\mathcal{D}}_{\mathrm{I}}=\frac{3}{32}\left(Ca_{\mathrm{core}}/\lambda \right){\beta }_{\mathrm{i}}^{4}{\left({\beta }_{\mathrm{i}}-{\beta }_{0}\right)}^{2}+3Ca_{\mathrm{il}}\left({\delta }_{0}/s_{0}\right)\ln^{2}\left(A_{\mathrm{i}}/A_{0}\right)$. Equations~(\ref{chi}) and (\ref{stageI}) close the problem for ${\beta }_{\mathrm{i}}$ and $\chi$, with $C_{\mathrm{sq}}$ as the only fitted constant. With $C_{\mathrm{sq}}=0.8$ (data constrain $0.6\lesssim C_{\mathrm{sq}}\lesssim 0.9$; SM, Sec. S8), the model reproduces ${\beta }_{\mathrm{i}}\approx 1.45-1.5$ and its weak growth with $We_{0}$ [Figs. 3(f) and 3(g)]. It underestimates $\chi$ by $\approx25\%$ because the front decelerates abruptly in the last $5\,ms$ [Fig. 2(a)], which the linear mean-velocity estimate does not capture. The energy budget [Fig. 5(a)] shows that the squeeze flow is the dominant sink, absorbing 60\%-65\% of the available energy, while $\approx30\%$ is stored as surface energy. The residual kinetic energy, core dissipation, and shell dissipation take only 2\%-4\%, 2\%-3\%, and $\approx1\%$, respectively.

\begin{figure*}
    \includegraphics[width=\textwidth]{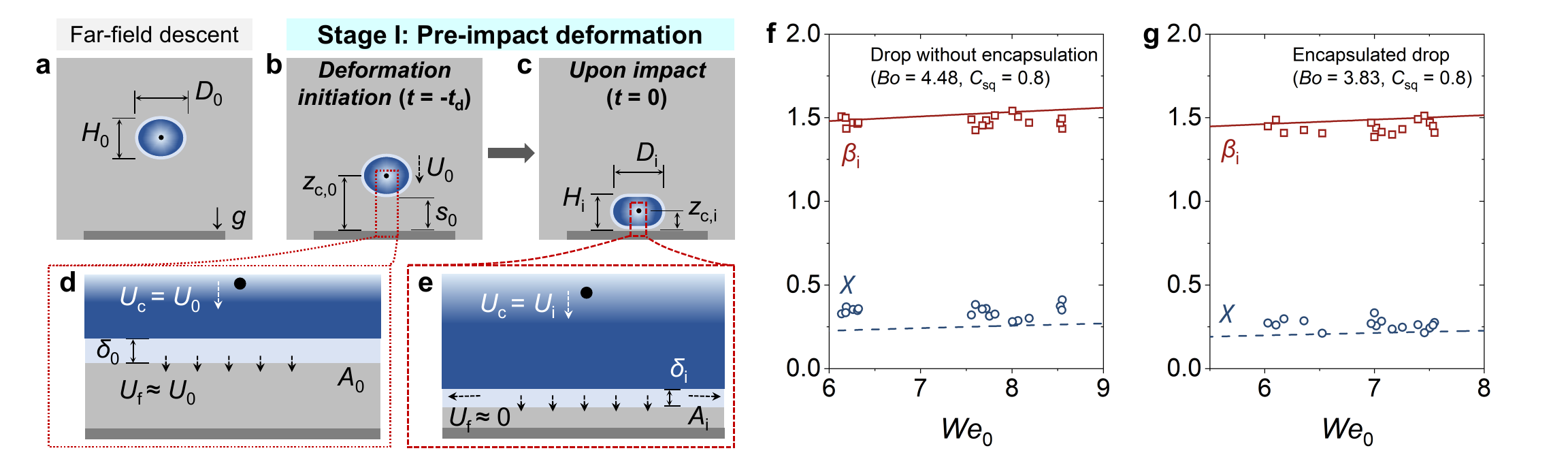}
    \caption{\label{fig3} Stage I: squeeze-flow pre-deformation. (a)-(c) Geometry from far-field descent through the onset of deformation ($t=-t_{\mathrm{d}}$) to impact ($t=0$). (d),(e) Close-ups: $U_{\mathrm{f}}\approx U_{\mathrm{c}}=U_{0}$ at onset; $U_{\mathrm{f}}\approx 0$ and $U_{\mathrm{c}}=U_{\mathrm{i}}$ at impact, while the shell thins from ${\delta }_{0}$ to ${\delta }_{\mathrm{i}}$ as its area grows. (f),(g) Measured (symbols) and predicted [Eqs. (\ref{chi}) and (\ref{stageI}), lines] ${\beta }_{\mathrm{i}}$ and $\chi$ for (f) bare ($Bo=4.48$, $\lambda=0.35$) and (g) AP1000-encapsulated ($Bo=3.83$, $\lambda=0.38$, ${\delta }_{0}/s_{0}=0.04$) Krytox 1506 drops; $C_{\mathrm{sq}}=0.8$.}
\end{figure*}

\textit{Stage II: wake-powered spreading}---After impact [Figs. 4(a)–4(d)], the centroid decelerates from $U_{\mathrm{i}}$ to 0 over $\left(H_{\mathrm{i}}-H_{\mathrm{m}}\right)/2$, so that the spreading lasts
\begin{equation}
  t_{\mathrm{s}}\simeq \frac{2D_{\mathrm{eq}}\left({\beta }_{\mathrm{i}}^{-2}-{\beta }_{\mathrm{m}}^{-2}\right)}{3U_{\mathrm{i}}}.
  \label{ts}
\end{equation}
The drop now holds only $K_{\mathrm{i}}={\chi }^{2}K_{0}$ of kinetic energy, which cannot pay for the observed spreading. We consider that the missing source is the trailing wake flow. A drop translating at Reynolds number $\approx{10}^{3}$ drags behind it a wake extending a few diameters \cite{20,50}. The wall halts the drop, but the trailing liquid continues at a velocity of $\approx U_{0}$ and impinges on the decelerated drop. We write the wake flow energy as $\Delta K_{\mathrm{w}}\simeq \frac{1}{2}C_{\mathrm{w}}{\rho }_{\mathrm{bath}}VU_{0}^{2}$, where $C_{\mathrm{w}}$ lumps the wake volume relative to the drop volume ($\approx5.4$ for a wake of length $2.5D_{\mathrm{eq}}$ \cite{50}) and the fraction of its energy transferred ($\approx0.8$), giving $C_{\mathrm{w}}\approx 4.3$. The energy balance,
\begin{equation}
  K_{\mathrm{i}}+W_{\mathrm{g},\mathrm{II}}+\Delta K_{\mathrm{w}}=\Delta E_{\Gamma,\mathrm{II}}+E_{\mathrm{core},\mathrm{II}}+E_{\mathrm{il},\mathrm{II}}+E_{\mathrm{lub}},
  \label{E_stageII}
\end{equation}
with core dissipation in biaxial extension \cite{49}, shell dissipation from Eq. (\ref{EilI}), and lubrication-film dissipation (SM, Sec. S7) then becomes, 
\begin{equation}
  S\left({\beta }_{\mathrm{m}}\right)-S\left({\beta }_{\mathrm{i}}\right)=\frac{We_{\mathrm{eff}}}{6}+\frac{Bo}{9}\left({\beta }_{\mathrm{i}}^{-2}-{\beta }_{\mathrm{m}}^{-2}\right)-{\mathcal{D}}_{\mathrm{II}},
  \label{stageII}
\end{equation}
\begin{equation}
  We_{\mathrm{eff}}={\chi}^{2}We_{0}+C_{\mathrm{w}}We_{\mathrm{b}},
  \label{Weeff}
\end{equation}
where $We_{\mathrm{b}}={\rho }_{\mathrm{bath}}U_{0}^{2}D_{\mathrm{eq}}/\Gamma$ is the Weber number of the ambient liquid and ${\mathcal{D}}_{\mathrm{II}}$ collects the normalized dissipations. Fed with ${\beta }_{\mathrm{i}}$ and $\chi$ from stage I, Eq. (\ref{stageII}) captures the magnitude and trend of measured ${\beta }_{\mathrm{m}}$ for drops [Figs. 4(e) and 4(f)], including the reduced spreading of the more viscous Krytox 1514 drop (SM, Sec. S9).

The effective Weber number $We_{\mathrm{eff}}$ in Eq. (\ref{Weeff}) combines the contributions of drop and wake inertia. Its two terms compare the residual drop inertia with that carried by the ambient liquid. The ratio $C_{\mathrm{w}}{\rho }_{\mathrm{bath}}/\left({\chi }^{2}{\rho }_{\mathrm{core}}\right)\approx 25-60$ shows that underliquid spreading is governed by $We_{\mathrm{b}}$ rather than $We_{\mathrm{0}}$. In the inviscid limit, Eq. (\ref{stageII}) simplifies to ${\beta }_{\mathrm{m}}\simeq {\left(C_{\mathrm{w}}We_{\mathrm{b}}/6\right)}^{1/2}$ for $We_{\mathrm{b}}\gg 1$, the classical capillary-inertial energy scaling \cite{3,5} with the drop density replaced by that of the surrounding liquid. In air, ${\rho }_{\mathrm{bath}}/{\rho }_{\mathrm{core}}\sim {10}^{-3}$, and Eq. (\ref{stageII}) recovers the closed budget of airborne impact. The scaling makes a falsifiable prediction: at fixed drop inertia, spreading should increase with the density of the ambient liquid, and decrease when the wake is suppressed.

\begin{figure*}
    \includegraphics[width=\textwidth]{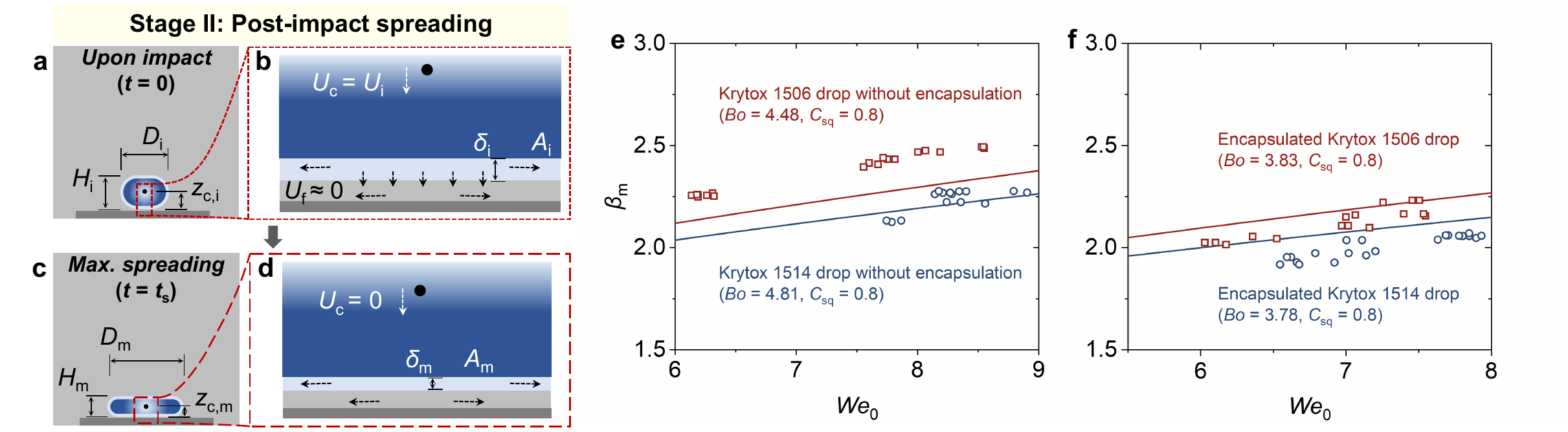}
    \caption{\label{fig4} Stage II: wake-powered spreading. (a)–(d) Geometry at impact ($t=0$) and at maximum spreading ($t=t_{\mathrm{s}}$), with close-ups of the shell and the water film. (e),(f) Measured (symbols) and predicted [Eq. (\ref{stageII}), lines] maximum spreading factor for (e) bare Krytox 1506 ($Bo=4.48$) and 1514 ($Bo=4.81$) drops and (f) AP1000-encapsulated Krytox 1506 ($Bo=3.83$) and 1514 ($Bo=3.78$) drops; $C_{\mathrm{sq}}=0.8$, $C_{\mathrm{w}}=4.3$, ${\delta }_{0}/s_{0}=0.04$.}
\end{figure*}

\textit{Energy partition and role of the shell}---The energy budgets of Fig. 5 make the two acts of the ambient liquid explicit. In stage I, the squeeze flow is the main destination at every $We_{0}$. In stage II, the wake supplies at least 90\% of the available energy over $2\leq We_{0}\leq 10$ and $1\leq Bo \leq8$ (SM, Sec. S10), whereas the residual kinetic energy and gravity together contribute less than 10\%. Surface energy remains the main destination in stage II (72\%-91\%), but its share falls as $We_{0}$ grows and viscous losses increase. The shell illustrates how dissipation depends on strain. Equation (\ref{EilI}) shows that shell dissipation scales with the square of the Hencky areal strain, $\ln\left(A_{2}/A_{1}\right)$. Stage I enlarges the surface area by only $\approx 20\%$, whereas stage II enlarges it by $\approx 75\%$, so the $\ln^{2}\left(A_{2}/A_{1}\right)$ increases ninefold between stages of comparable duration. Accordingly, a shell about $60\,\mu m$ thick dissipates $\approx1\%$ of the energy before impact but $\approx 6\%$ during spreading, and its share grows to 30\% for thicker or more viscous shells (SM, Sec. S10). Before impact, encapsulation therefore acts almost entirely by lowering $\Gamma$. During spreading, the shell, whereas, becomes an appreciable viscous brake. 

\textit{Conclusions}---Underliquid impact of drops larger than the capillary length proceeds in two acts played by the ambient liquid. The squeeze flow in the draining gap first flattens the drop and removes most of its kinetic energy before its front stops. The wake then returns energy to the decelerated drop and drives it to spread on the water lubrication film. The surface energy increase therefore exceeds the drop’s own mechanical energy, and the maximum spreading is governed by the Weber number of the ambient liquid rather than that of the drop. Because the lubrication film prevents contact, spreading is insensitive to wettability. A thin shell modifies it through its interfacial tension and, once stretched, through biaxial-extension dissipation. These results provide a framework for controlling the deposition and transfer of encapsulated drops in liquid environments. They also identify the trailing wake as a quantity to be measured directly. Time-resolved flow-field measurements would calibrate $C_{\mathrm{w}}$ and $C_{\mathrm{sq}}$ and test the predicted dependence of spreading on ambient properties. 

\begin{figure*}
    \includegraphics[width=\textwidth]{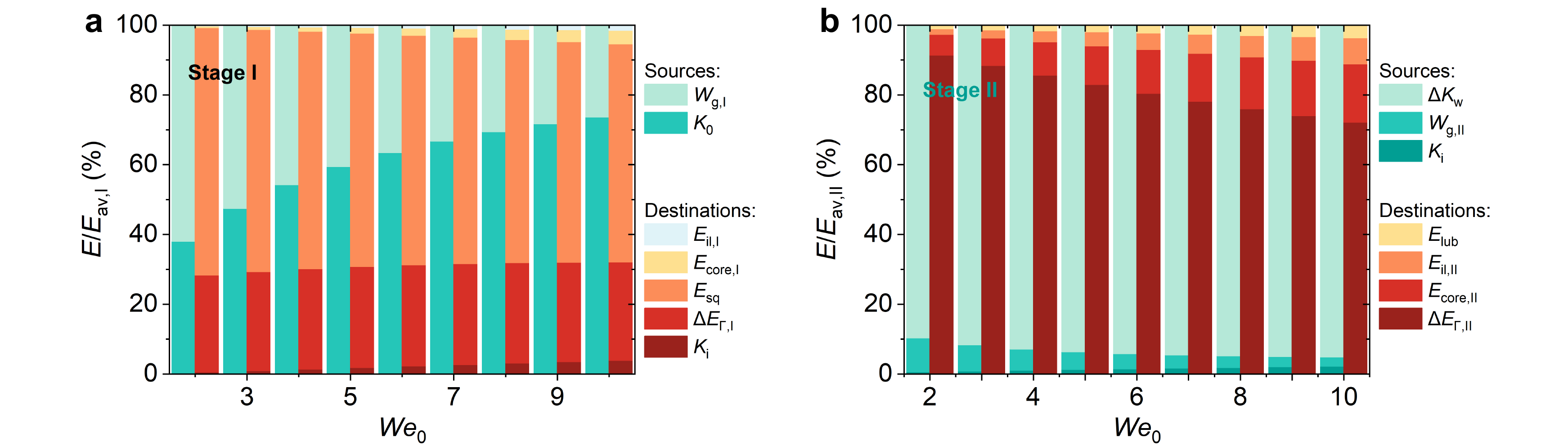}
    \caption{\label{fig5} Predicted energy budgets of encapsulated drops ($Bo=3.83$) versus $We_{0}$ in (a) stage I and (b) stage II. In each pair, the left column shows the sources and the right column the destinations, normalized by $E_{\mathrm{av},\mathrm{I}}=K_{0}+W_{\mathrm{g},\mathrm{I}}$ and $E_{\mathrm{av},\mathrm{II}}=K_{\mathrm{i}}+W_{\mathrm{g},\mathrm{II}}+\Delta K_{\mathrm{w}}$. Parameters as in Figs. 3 and 4.}
\end{figure*}

\textit{Acknowledgments}---This research was supported by the Natural Sciences and Engineering Research Council of Canada (NSERC) through a Discovery Grant to S. K. M. (RGPIN-2024-03729) and by a startup grant from the University of Waterloo. T. Z. thanks Arnav Banerjee for assistance with surface wettability tests. The authors used Claude (Anthropic, v4.8 and v5.5) to assist in checking the derivations and in editing the manuscript; the authors reviewed all content and take full responsibility for it.

\textit{Author contributions}---S. K. M. conceived and supervised the research. T. Z. performed the experiments and analysis, and developed the model. Both authors wrote the manuscript.

\textit{Data availability}---The data that support the findings of this Letter are available from the corresponding author upon reasonable request. 

\nocite{39,40,41,42,43,44,45,46,47,48}
\bibliography{prl_ref}

\end{document}